\documentclass[prb,twocolumn,aps,superscriptaddress,floatfix]{revtex4}
\usepackage{amsmath}
\usepackage{amssymb}
\usepackage{amsfonts}
\usepackage{bm}
\usepackage[dvips]{graphicx}
\usepackage{color}
\usepackage{ulem}
\usepackage{verbatim}
\DeclareGraphicsExtensions{.eps,.pdf,.png}
\graphicspath{{pictures/}}
\usepackage{epstopdf}
\let\emph\relax 
\DeclareTextFontCommand{\emph}{\it}

\begin{document}

\title{Dynamical spin susceptibility of spin-valley half-metal}

\author{D.A. Khokhlov}
\affiliation{Moscow Institute of Physics and Technology, Institutsky lane
9, Dolgoprudny, Moscow region, 141700}
\affiliation{Dukhov Research Institute of Automatics, Moscow, 127055}
\affiliation{National Research University Higher School of Economics}

\author{A.L. Rakhmanov}
\affiliation{Moscow Institute of Physics and Technology, Institutsky lane
9, Dolgoprudny, Moscow region, 141700}
\affiliation{Dukhov Research Institute of Automatics, Moscow, 127055}
\affiliation{Institute for Theoretical and Applied Electrodynamics, Russian
Academy of Sciences, Moscow, 125412 Russia}

\author{A.V. Rozhkov}
\affiliation{Moscow Institute of Physics and Technology, Institutsky lane
9, Dolgoprudny, Moscow region, 141700}
\affiliation{Institute for Theoretical and Applied Electrodynamics, Russian
Academy of Sciences, Moscow, 125412 Russia}

\author{A.O. Sboychakov}
\affiliation{Institute for Theoretical and Applied Electrodynamics, Russian
Academy of Sciences, Moscow, 125412 Russia}

\begin{abstract}
A few years ago we predicted theoretically that in systems with nesting of
the Fermi surface the spin-valley half-metal has lower energy than the spin
density wave state. In this paper we suggest a possible way to distinguish
these phases experimentally. We
calculate dynamical spin susceptibility tensor for both states in the
framework of the Kubo formalism. Discussed phases have different numbers of
the bands: four bands in the spin-valley half-metal and only two bands in
the spin density wave. Therefore, their susceptibilities, as functions of
frequency, have different number of peaks. Besides, the spin-valley
half-metal does not have rotational symmetry, thus, in general the
off-diagonal components of susceptibility tensor are non-zero. The spin
density wave obeys robust rotational symmetry and off-diagonal components of
the susceptibility tensor are zero. These characteristic features can be
observed in experiments with inelastic neutron scattering.
\end{abstract}
\date{\today}

\maketitle
\section{Introduction}
The concept of Fermi surface nesting plays an important role in the studies
of systems with more than one electronic
band~\cite{rice1970,ourPRB_phasepAFM2017,ourPRL_half-met2017,
Chuang1509,PhysRevB.86.020507,PhysRevB.82.020510,PhysRevB.94.085106,
mosoyan_aa_graphit2018,ourPRB_half-met2018,Nandkishore2012,
our_PRB_magfield_imp_nest,gonzalez_kohn-lutt_twist2019,
twist_graph_bias_gap_prl2018,nesting_review2017,
aa_graph2014,aa_graph2013,phasep_pnics2013,q1d2009,q1d2009_2,q1d2003,
hirschfeld_review2011,fernandez_pnic_so5_2010,gruner_book}.
In the presence of nesting electronic liquid becomes unstable and ordered
state emerges. The ordered phases associated with the nesting are
commensurate and incommensurate spin density
wave~\cite{rice1970} (SDW),
charge density
wave~\cite{PhysRevB.77.165135},
inhomogeneous
antiferromagnetism~\cite{ourPRB_phasepAFM2017},
as well as others. Recently we argued that a half-metallic and the
so-called spin-valley half-metal (SVHM) may be stabilized in a system with
nesting~\cite{ourPRB_half-met2018,Nandkishore2012}.
Half-metals~\cite{first_half_met1983,half_met_review2008,
sc_half_met_eshrig2015}
are well known for a several decades. This class of metals is characterized
by perfect spin polarization of the charge carriers on the Fermi surface.
Consequently, electric current in a half-metal carries not only charge but
spin as well. The latter property is of interest for applications in
spintronics. By the same token, the electronic states at the Fermi surface
of the spin-valley half-metal are perfectly polarized with respect to the
so-called spin-valley index. Thus, the SVHM can conduct a spin-valley
polarized current. The spin-valley polarized currents are of interest for
applications~\cite{Zhonge1603113}.
In addition, the SVHM state does not require strong electron-electron
interaction for its stability. Consequently, the SVHM can be realized
without transition metals in its chemical composition, which makes such
materials applicable in biocompatible devices.

Theoretically, the SVHM state was identified in a model with perfect
nesting and weak electron-electron interaction. When the perfect nesting is
partially destroyed by doping, parental insulating SDW is replaced by the
SVHM
phase~\cite{ourPRL_half-met2017}:
calculations within the framework of such a minimal model show that the
SVHM phase has lower free energy than the SDW and the paramagnetic phase,
at least, for low
temperatures~\cite{ourPRL_half-met2017}.
Depending on details, the SVHM may be either commensurate or
incommensurate~\cite{ourPRL_half-met2017,ourPRB_half-met2018}.
It competes against several related phases, such as (i)~commensurate SDW,
(ii)~incommensurate SDW, (iii)~inhomogeneous SDW.

As shown in
Refs.~\onlinecite{ourPRL_half-met2017,ourPRB_half-met2018},
the SVHM has a number of distinctive features which separate it from
phases~(i-iii). For example, magnetic structure of the SVHM state possesses
a magnetic helical component, which superimposes on the purely collinear SDW
order. Beside this, the symmetry between single-electron states with
different spin-valley index is preserved in the ordered states (i-iii), but
is lifted in the SVHM. As a result, the number of non-degenerate
single-electron bands in the SVHM is two times bigger than in the SDW.

However, the SVHM has not been observed experimentally yet. In this paper,
we discuss a possibility of detecting the SVHM phase using inelastic
neutron scattering. For many materials, neutron scattering has been
successfully applied to investigate their magnetic and superconducting
properties. Comparing observed spectrum with a theoretical prediction,
useful pieces of information can be
obtained~\cite{PhysRevB.78.140509,PhysRevB.77.165135,
NatureSupercondNesting2012,fine_neutrons2007,fine_review2010,
egami_physC2010,Lychkovskiy2017SpinES,PhysRevB.60.3643,
PhysRevB.78.052508,Mazin_1995,PhysRevLett.108.117001}.
It is natural to expect that such an experimental tool can play important
role in search for the SVHM order.

Below we propose a method to discriminate between the SVHM and the SDW
phases. It relies on the fact that the SVHM has at least four
non-degenerate bands close to the Fermi level, whereas the SDW has two
doubly degenerate bands. Consequently, in these thermodynamic phases, the
electronic contributions to the neutron cross-section are non-identical,
and phase-specific features in the neutron spectrum can be used for
identification of an ordered state in a candidate material.

Technically, the electronic contribution to the neutron cross-section is
described by the dynamical spin susceptibility tensor. Using Kubo formalism
within the framework of the minimal model of
Refs.~\onlinecite{ourPRL_half-met2017,ourPRB_half-met2018}
we determine this tensor for the commensurate SVHM and SDW phases. Our
calculations demonstrate that the neutron scattering spectrum of the SVHM
has three high-intensity peaks, in contrast to the SDW, whose spectrum has
only one pronounced peak. We also discuss other features of the spin
susceptibility tensor that may be used to identify the SVHM phase with on a
neutron scattering experiment.

This paper is organized as follows. In Sec.~\ref{sec::Model} we briefly describe two band
model used in calculations. In Sec.~\ref{sec::sdw} and~\ref{sec::svhm} we calculate the
susceptibility tensor for the SDW and the SVHM respectively. Summary and
conclusions are in Sec.~\ref{sec::discussion}.
\section{Model}
\label{sec::Model}
\subsection{SVHM and SDW phases}
\begin{figure*}
 	\center{\includegraphics[width=1\linewidth]{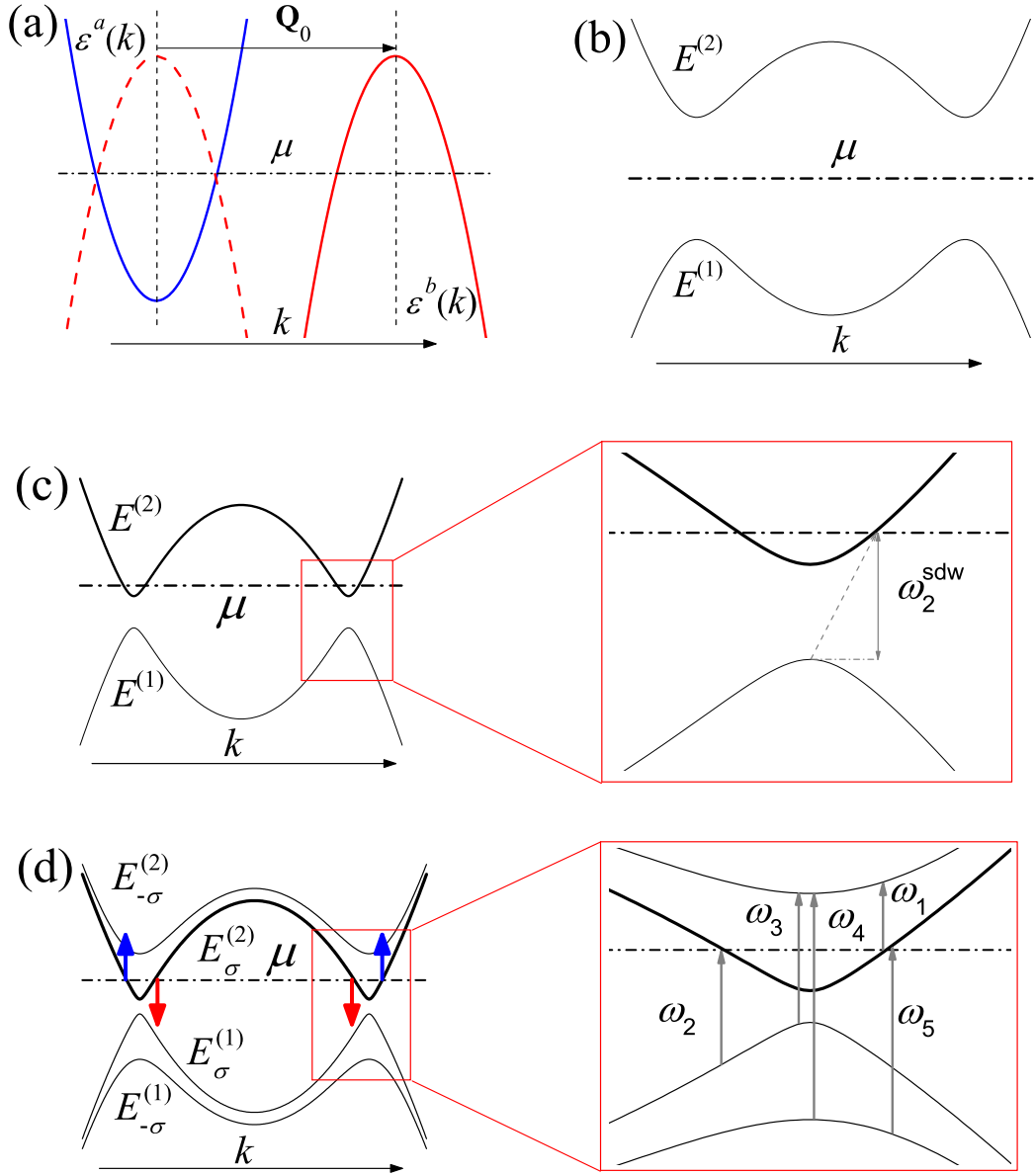}}
	\caption{
The energy of single-electron and quasiparticle bands (vertical axis)
versus momentum (horizontal axis) for different order parameters and doping
values. (a)~Bare single-particle bands. The (blue) solid line represents
the electron dispersion (band $a$), the (red) solid line is the hole
dispersion (band $b$). The translated hole band is shown by dashed line.
Arrow
${\bf Q}_0$
corresponds to the nesting vector. The dashed-dotted horizontal line is the
Fermi energy of the undoped state, when the nesting is perfect.
(b)-(d)~Band structure when the effects of weak electron-electron repulsion
are taken into account.
(b)~The band structure of the undoped SDW state. The chemical potential
level (dash-dotted line) lies in the spectral gap separating filled band
$E^{(1)}$
and empty band
$E^{(2)}$.
Both bands are doubly degenerate.
(c)~The band structure of the doped SDW state. The bands are shown by solid
lines. The chemical potential crosses the upper band
$E^{(2)}$
which becomes partially filled. The vertical arrow in the insets indicates
quasiparticle interband transition at the threshold frequency
$\omega^{\rm sdw}_2$,
see
Eq.~(\ref{Eq::omega_2_sdw}).
(d)~The quasiparticle band structure of the SVHM phase. Double degeneracy
present in the SDW state is lifted, and all four bands are distinct.
The Fermi surface is polarized with respect to the spin-valley index. Up
and down bold arrows illustrate this polarization. Up-arrows (blue) show
spin-up electrons from the valley $a$, down-arrows (red) show spin-down
electrons from the valley $b$. The vertical arrows in the inset illustrate
minimum energies which are required to open a specific interband scattering
channel. The frequencies
$\omega_{1}$,
$\ldots$,
$\omega_{5}$
are defined by
Eq.~(\ref{Eq::svhm_omega_n}).
 	\label{Fig::bands&peaks}
 	}
\end{figure*}

We start with the outline of the basic structure of the minimal
model~\cite{ourPRL_half-met2017}
which hosts the SVHM as one of its possible ground states. The model
Hamiltonian has two single-electron bands, or valleys, which are referred
to as $a$ and $b$. If we neglect the electron-electron repulsion, their
band dispersions are assumed to be parabolic
(Fig.~\ref{Fig::bands&peaks}a)
\begin{eqnarray}
	\label{Eq::disperse_no_order}
	\epsilon_{a}(\mathbf{k})=\frac{k^2-k_F^2}{2m_a}-\mu=\xi_k^a-\mu,\\ \nonumber
	\epsilon_{b}(\mathbf{k}+\mathbf{Q}_0)=\frac{-k^2+k_F^2}{2m_b}-\mu=-\xi_k^b-\mu.
\end{eqnarray}
We use system of units, where
$\hbar=1$.
Here
$\mu$
is the chemical potential. When
$\mu=0$,
the Fermi surface of the valley $a$, after translation by the nesting
vector
$\mathbf{Q}_0$,
exactly matches the Fermi surface of the valley $b$. Such a property of the
band structure is called a perfect nesting. It is convenient to measure
doping relative to
$\mu=0$
state treating the latter state as undoped. Momentum
$k_F=\sqrt{2m\epsilon_F}$
is a radius of the Fermi sphere for the both $a$ and $b$ valleys at the
perfect nesting. In addition, we introduce the Fermi velocity
$v_F=k_F/m$.
Unless the opposite is stated, we assume that effective masses
$m_a$
and
$m_b$
are equal to each other. In such a case, the subscript may be dropped, both masses can be denoted by symbol $m$, and $\xi_k^a=\xi_k^b=\xi_k$. Each band has a density of states
$N_F=mk_F/(2\pi^2)$
at the Fermi level.

Next, we take into account a weak electron-electron repulsion, which we
assume to be short-range. The part of the interaction which is responsible
for the magnetic ordering is:
\begin{eqnarray}
	\label{Eq::hamiltonian_int}
H_{\text{int}}=g\sum_{\mathbf{k}\mathbf{k}'\sigma}a^{\dagger}_{\sigma\mathbf{k}}a^{\phantom{\dagger}}_{\sigma\mathbf{k}} b^{\dagger}_{\overline{\sigma}\mathbf{k}'}b^{\phantom{\dagger}}_{\overline{\sigma}\mathbf{k}'}\,,
\end{eqnarray}
where operator
$a_{\sigma\mathbf{k}}$
(operator
$b_{\sigma\mathbf{k}}$
represents annihilation operator of an electron with the spin $\sigma$ in
the valley $a$ (valley $b$) at the wave vector
$\mathbf{k}$.
For the operators
$a_{\sigma\mathbf{k}}$
and
$b_{\sigma\mathbf{k}}$,
the wave vector
$\mathbf{k}$
is measured from the center of the corresponding valley. Notation
$\overline{\sigma}$
means
$-\sigma$.
The interaction constant $g$ is assumed to be small:
$gN_F\ll1$.
We simplify the Hamiltonian via mean field approach. There are two order
parameters labeled by
$\sigma=\pm 1$

\begin{eqnarray}
	\label{Eq::delta_s_def}
	\Delta_{\sigma}=\frac{g}{V}\sum_{\mathbf{k}}\langle a^{\dagger}_{\sigma\mathbf{k}}b^{\phantom{\dagger}}_{\overline{\sigma}\mathbf{k}}\rangle,
\end{eqnarray}
where $V$ is the volume of the system.

To diagonalize the Hamiltonian we perform the Bogolyubov transformation. The
obtained quasiparticle spectrum consists of four bands:
\begin{eqnarray}
	\label{Eq::dispers_svhm}
	E_{\sigma\mathbf{k}}^{(s)}=\pm\sqrt{\xi_k^2+\Delta_{\sigma}^2},
\end{eqnarray}
where each band is labeled by two indexes: $\sigma$ and
$s=1,2$.
Here
$s=1$
corresponds to sign ``-" and
$s=2$
to sign ``+". To obtain
$\Delta_{\sigma}$
we should minimize the total energy of the system. After the minimization
at zero doping one may see that the order parameters does not depend on the
index $\sigma$, thus, the quasiparticle bands are double degenerate.
Consequently,
$\mu=0$
state posses SDW order with static spin polarization
\begin{eqnarray}
	\label{Eq::av_spin_sdw}
	\langle S_x(\mathbf{r})\rangle=\frac{\Delta_{\sigma}+\Delta_{\overline{\sigma}}}{g}\cos(\mathbf{rQ}_0), \langle S_y(\mathbf{r})\rangle=\langle S_z(\mathbf{r})\rangle=0.
\end{eqnarray}
Band structure of the SDW is schematically shown in
Fig.~\ref{Fig::bands&peaks}b.

Now let us study the effect of the doping on the SDW state. In many
papers~\cite{ourPRB_phasepAFM2017,rice1970,our_PRB_magfield_imp_nest}
the energy minimization was performed under the following constraint
\begin{eqnarray}
	\label{Eq::order_par_sdw}
	\Delta_{\sigma}=\Delta_{\overline{\sigma}}=\Delta.
\end{eqnarray}
Minimizing the total energy under the
condition~(\ref{Eq::order_par_sdw})
we find $\Delta$ as a  function of doping. The resultant state is the SDW
metal~\cite{ourPRB_phasepAFM2017,rice1970}.
Quasiparticle bands at non-zero doping are shown in
Fig.~\ref{Fig::bands&peaks}c.
They remain doubly degenerate, since we employ
restriction~(\ref{Eq::order_par_sdw}).

However, in a more general case the
condition~(\ref{Eq::order_par_sdw})
can be discarded. Instead, the total energy is minimized as a function of
two variables
$\Delta_{\sigma}$
and
$\Delta_{\overline{\sigma}}$.
The latter minimization is simplified by the fact that the mean field
Hamiltonian split into a sum of two decoupled terms, each describing a
particular sector of single-particle states. The first term represents
(i)~electrons from the valley $a$ with the spin $\sigma$ and (ii)~electrons
from the valley $b$ with the spin
$\overline{\sigma}$.
These quasiparticle states form sector $\sigma$. The order in sector
$\sigma$ is characterized by
$\Delta_\sigma$.

The second term of the mean field Hamiltonian represents electrons from the
valley $a$ with spin
$\overline{\sigma}$
and from the valley $b$ with spin $\sigma$. Such states constitute sector
$\overline{\sigma}$.
Parameter
$\Delta_{\overline{\sigma}}$
describes order in sector
$\overline{\sigma}$.

Doping is not required to distribute equally between the sectors. Moreover,
it was observed in
Refs.~\onlinecite{our_PRB_magfield_imp_nest,ourPRL_half-met2017}
that the total energy of the doped system is the lowest when all doped
electrons enter a single sector keeping the other sector completely empty.
For definiteness, we will assume below that all doping $x$ accumulates in
the sector $\sigma$.

Since our minimization is not constrained by
condition~(\ref{Eq::order_par_sdw})
we obtain different order parameters in each sectors:
\begin{eqnarray}
	\label{Eq::order_par_svhm}
	\Delta_{\sigma}=\Delta_0\sqrt{1-\frac{x}{N_F \Delta_0}}, \;\; \Delta_{\overline{\sigma}}=\Delta_0.
\end{eqnarray}
where
$\Delta_0$
is the order parameter at zero doping in the SDW phase.  In contrast with
the SDW phase, the SVHM has two non-zero spin projections:
\begin{eqnarray}
	\label{Eq::av_spin_svhm}
	\langle S_x(\mathbf{r})\rangle=\frac{\Delta_{\sigma}+\Delta_{\overline{\sigma}}}{g}\cos(\mathbf{rQ}_0),
\\ \nonumber
\langle  S_y(\mathbf{r})\rangle
=
\frac{\Delta_{\sigma}-\Delta_{\overline{\sigma}}}{g}
\sin(\mathbf{rQ}_0).
\end{eqnarray}
Four bands which are defined in
Eq.~(\ref{Eq::dispers_svhm})
are no longer degenerate, and the Fermi surface appears.
We may define the spin-valley index~\cite{ourPRL_half-met2017}
$S_{v}$
as follows
\begin{eqnarray}
S_{v}=1 \leftrightarrow\text{electronic states from sector} \;\sigma,
\\ \nonumber
S_{v}=-1 \leftrightarrow\text{electronic states from sector}
\;\overline{\sigma}.
\end{eqnarray}
In
Ref.~[\onlinecite{ourPRL_half-met2017}]
it was noted that, when all the doping enters a single sector, keeping the
other sector empty, the Fermi surface states are polarized in the
spin-valley space, see
Fig.~\ref{Fig::bands&peaks}d.
This fact can be trivially verified since the partially filled band is
composed entirely from the electronic states belonging to sector $\sigma$.
Following
Ref.~[\onlinecite{ourPRL_half-met2017}],
we call this phase the spin-valley half-metal.

\subsection{The dynamical spin susceptibility tensor}
\label{subsec::susceptibility}
Here we discuss the dynamical spin susceptibility tensor in the context of
our problem. The Fourier components of the spin projection on the
$\beta$-axis is:
\begin{eqnarray}
	\label{Eq::spin_f_and_s}
S_{\beta}^{\rm s}(\mathbf{q},t)
=\!\!
\sum_{\mathbf{k}\mu\nu}\!\!
	\sigma^{\beta}_{\mu\nu}\!\!
	\left[
		a^{\dagger}_{\mu\mathbf{k}+\mathbf{q}}\!(t)\,
		a_{\nu\mathbf{k}}^{\vphantom{\dagger}} \!(t)
		\!+\!
		b^{\dagger}_{\mu\mathbf{k}+\mathbf{q}}\!(t)\,
		b_{\nu\mathbf{k}}^{\vphantom{\dagger}} \!(t)
	\right]\!,
\\
\nonumber
S_{\beta}^{\rm f}(\mathbf{q}+\mathbf{Q}_0,t)
=
\sum_{\mathbf{k}\mu\nu}
	\sigma^{\beta}_{\mu\nu}
	a^{\dagger}_{\mu\mathbf{k}+\mathbf{q}}(t)
	b_{\nu\mathbf{k}}^{\vphantom{\dagger}} (t)\,,
\\
\nonumber
S_{\beta}^{\rm f}(\mathbf{q}-\mathbf{Q}_0,t)
=
\sum_{\mathbf{k}\mu\nu}
	\sigma^{\beta}_{\mu\nu}
	b^{\dagger}_{\mu\mathbf{k}+\mathbf{q}}(t)
	a_{\nu\mathbf{k}}^{\vphantom{\dagger}} (t)\,.
\end{eqnarray}
Here
$\sigma^{\beta}_{\mu\nu}$
is
$(\mu,\nu)$
matrix element of a Pauli matrix and
$\beta=x,y\;\text{or}\;z$.
The superscript `s' (superscript `f') stands for `slow' (`fast'). The slow
term
$S_{\beta}^{\rm s}(\mathbf{q},t)$
oscillates in the real space with the wave vector
$\mathbf{q}$,
which we restrict to be in the range of
$ q \lesssim \Delta_0/v_F\ll k_F \sim |\mathbf{Q}_0|$,
where
$q = |{\bf q}|$.
This term contains only products of operators from one valley. In other
words, it is diagonal in valley index.
The fast terms
$S_{\beta}^{\rm f}(\mathbf{q}\pm\mathbf{Q}_0,t)$
oscillate in the real space with the wave vectors
$\mathbf{q}\pm\mathbf{Q}_0$.
Unlike
$S^{\rm s}$,
operators
$S^{\rm f}$
mix states from different valleys. Indeed, as one can see from their
definition, each
$S^{\rm f}$
is a sum of terms that themselves are products of two single-electron
operators, one single-electron operator from valley $a$ and another one is
from valley $b$.

The susceptibility is defined in the Kubo
formalism~\cite{mahan2013many}
\begin{eqnarray}
	\label{Eq::kubo_four_Q}
	\chi_{\alpha\beta}^{\rm s(f)}(\mathbf{Q},\omega)\!=\!i\!\!\int_0^{\infty}\!\!\!\!\!\!\langle[ S_{\alpha}^{\rm s(f)}(\mathbf{Q},t);S_{\beta}^{\rm s(f)}(-\mathbf{Q},0)] \rangle e^{i\omega t} dt,
\end{eqnarray}
where
$[A;B]=AB-BA$.
Symbol
$\langle...\rangle$
denotes averaging with respect to a ground state of a studied phase. We
assume that
$\mathbf{Q}=\mathbf{q}$
for the slow term and
$\mathbf{Q}=\mathbf{q}\pm\mathbf{Q}_0$
for the fast terms. Besides
$\chi_{\alpha\beta}^{\rm s(f)}$
defined by
Eq.~(\ref{Eq::kubo_four_Q}),
it is possible to introduce the cross-terms, describing correlation
functions between slow and fast spin densities
$\chi^{\rm sf} \sim \langle S^{\rm s} S^{\rm f} \rangle$.
However, these quantities do not contribute to the neutron cross-section
and we do not consider them.

Further, as we want to describe neutron scattering, we focus only on the
part of the tensor which corresponds to the energy conservation law
\begin{eqnarray}
	\label{Eq::kubo_four_Q_tilde}
	\tilde{\chi}_{\alpha\beta}^{\rm s(f)}(\mathbf{Q},\omega)\!=\!i\!\!\int_{-\infty}^{\infty}\!\!\!\!\!\langle[ S_{\alpha}^{\rm s(f)}(\mathbf{Q},t);S_{\beta}^{\rm s(f)}(-\mathbf{Q},0)] \rangle e^{i\omega t} dt.
\end{eqnarray}
Here integration over all real $t$ gives Dirac
$\delta$-function
which ensures energy conservation in the scattering processes. Related
result in a less general case was discussed in 
Ref.~\onlinecite{furrer2009neutron} [see Eq.~(2.46) in section~2.6 for more details]. In our numerical calculation, the Dirac $\delta$-function
is approximated by a rectangular function. This function equals to
$10^3\Delta_0^{-1}$
over finite support, whose width is
$10^{-3}\Delta_0$
(as required by the definition of the Dirac function, the area under our
rectangular function is equal to unity).

Direct calculations show that all diagonal components
$\tilde{\chi}^{\rm f,s}_{\alpha\alpha}$
and off-diagonal components
$\tilde{\chi}^{\rm f,s}_{xy,yx}$
may be non-zero.
All other components of the dynamical susceptibility vanish within the
framework of the discussed model.
To calculate
$\tilde{\chi}_{\alpha\beta}^{\rm f,s}(\mathbf{Q},\omega)$
we substitute
Eq.~(\ref{Eq::spin_f_and_s})
into
formula~(\ref{Eq::kubo_four_Q_tilde})
and apply the Wick theorem with respect to the studied state. The results
of these calculations are presented below.

\section{Spin susceptibility of the SDW phase}
\label{sec::sdw}

\subsection{Structure of the susceptibility tensor in the SDW phase}
\label{subsec::gen_remarks}

We would like to start with several general statements about the
susceptibility tensor of the SDW phase. In the commensurate SDW state the
slow term
$\tilde{\chi}_{\alpha\beta}^{\rm s}(\mathbf{Q},\omega)$
of the susceptibility tensor does not depend on direction of the wave vector
$\mathbf{Q}=\mathbf{q}$,
and the fast term
$\tilde{\chi}_{\alpha\beta}^{\rm f}(\mathbf{Q},\omega)$
of the tensor does not depend on direction of the vector
$\mathbf{Q}-\mathbf{Q}_0=\mathbf{q}$.
The same property holds true in the commensurate SVHM phase discussed in
Sec.~\ref{sec::svhm}.
Obviously, this circumstance significantly simplifies the presentation and
analysis of our results.

Since the average local spin in the SDW is directed along the $x$-axis, the
magnetic structure of the phase possesses a rotational symmetry around
$x$-axis. Therefore, we obtain
\begin{eqnarray}
	\label{Eq::sdw_sus_relation_1}
	\tilde{\chi}_{\bot}^{\rm s}(\mathbf{q},\omega)
		=
		\tilde{\chi}_{yy}^{\rm s}(\mathbf{q},\omega)
		=
		\tilde{\chi}_{zz}^{\rm s}(\mathbf{q},\omega)
		\neq
		\tilde{\chi}_{xx}^{\rm s}(\mathbf{q},\omega),
\end{eqnarray}
where symbol
$\tilde{\chi}_{\bot}^{\rm s}$
is defined as
$\tilde{\chi}_{\bot}^{\rm s} = \tilde{\chi}_{zz,yy}^{\rm s}$.
Susceptibility
$\tilde{\chi}_{\bot}^{\rm f}$
is defined similarly.

As for the off-diagonal components,
$\tilde \chi_{xz}^{\rm s,f}$
and
$\tilde \chi_{yz}^{\rm s,f}$
vanish, as discussed in
subsection~\ref{subsec::susceptibility}.
The presence of the rotational symmetry with respect to $x$-axis implies
nullification of other off-diagonal components as well. Indeed, any
rotation around $x$-axis preserves the susceptibility tensor. At the same
time, after rotation on angle equal to $\pi$, components
$\tilde{\chi}_{xy,yx}^{\rm s,f}(\mathbf{Q},\omega)$
must change sign. Therefore, we conclude that
\begin{eqnarray}
	\label{Eq::sdw_sus_relation_2}
	\tilde{\chi}_{xy}^{\rm s,f}(\mathbf{Q},\omega)=\tilde{\chi}_{yx}^{\rm s,f}(\mathbf{Q},\omega)=0.
\end{eqnarray}
Since
relations~(\ref{Eq::sdw_sus_relation_1})
and~(\ref{Eq::sdw_sus_relation_2})
are conditioned by the rotational symmetry of the SDW order parameter, they
remain valid even when
$m_a \ne m_b$.
In other words, asymmetry between the electrons and holes does not
destroy~(\ref{Eq::sdw_sus_relation_1})
and~(\ref{Eq::sdw_sus_relation_2})
in the SDW phase.

Diagonal components of the fast part of the of the susceptibility tensor obey yet
another relation
\begin{eqnarray}
	\label{Eq::sdw_sus_relation_3}
	\tilde{\chi}_{\bot}^{\rm f}(\mathbf{q}\pm\mathbf{Q}_0,\omega)=\tilde{\chi}_{xx}^{\rm f}(\mathbf{q}\pm\mathbf{Q}_0,\omega).
\end{eqnarray}
It can be derived by substituting
Eq.~(\ref{Eq::spin_f_and_s})
into
Eq.~(\ref{Eq::kubo_four_Q_tilde}).
(Let us remark that
Eqs.~(\ref{Eq::sdw_sus_relation_1})
and~(\ref{Eq::sdw_sus_relation_2})
can be derived by the same substitution as well, without use of symmetry.)
\begin{figure*}
	\center{\includegraphics[width=1\linewidth]{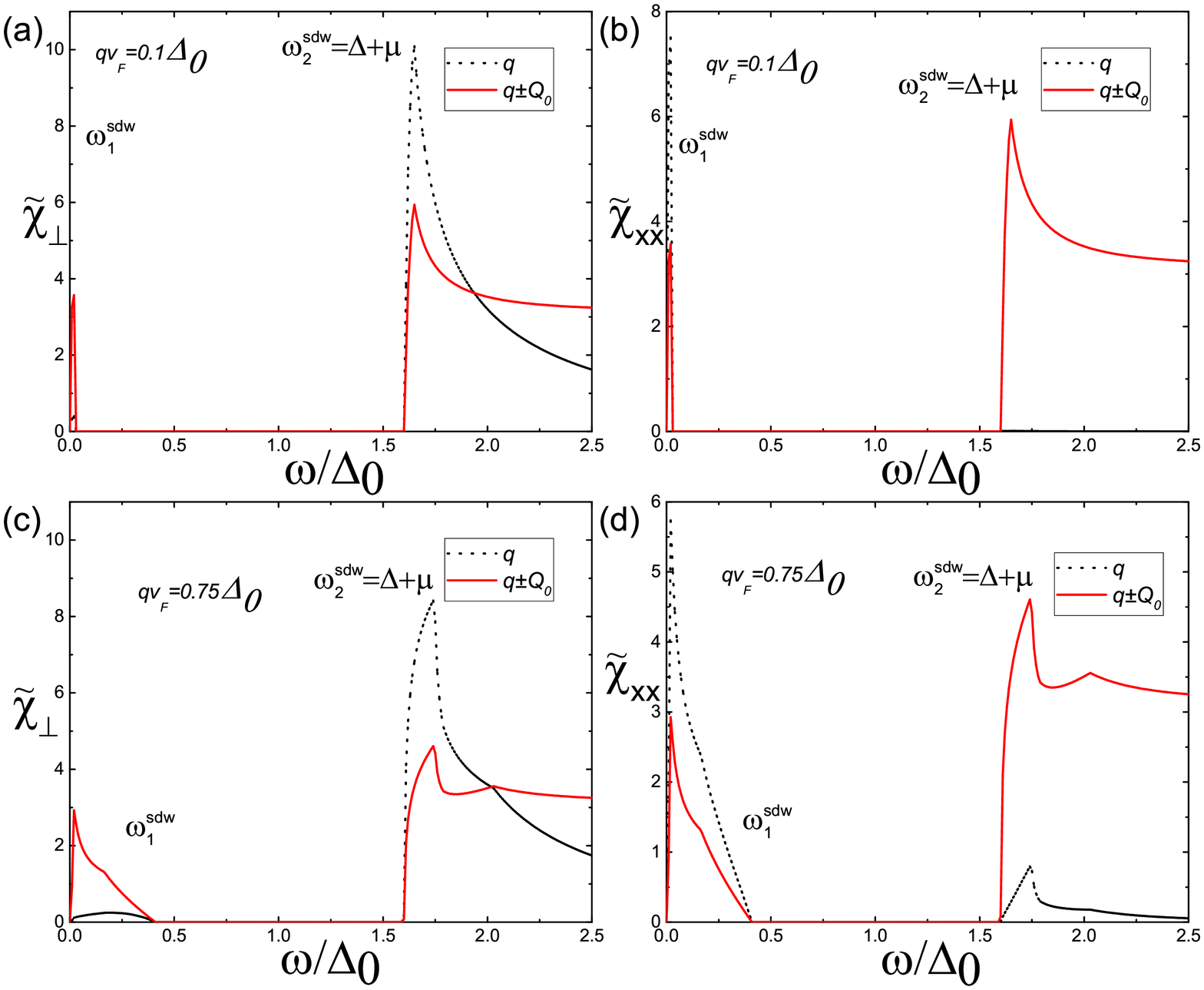}}
	\caption{The diagonal components of the susceptibility tensor
$\tilde{\chi}_{\alpha\beta}^{\rm s,f}(\mathbf{Q},\omega)$
at fixed momentum
$\mathbf{Q}$
versus frequency $\omega$ in the SDW phase. Dot curves show the slow part
of the susceptibility calculated for
$\mathbf{Q}=\mathbf{q}$;
solid curves show the fast part of the susceptibility calculated for
$\mathbf{Q}=\mathbf{q}\pm\mathbf{Q}_0$.
Panels~(a) and~(c) present
$\tilde{\chi}_{\bot}(\mathbf{Q},\omega)$,
panels (b),(d) present
$\tilde{\chi}_{xx}(\mathbf{Q},\omega)$.
The data in panels~(a) and~(b) is plotted for
$q=0.1\Delta_0/v_F$.
The data in panels~(c) and~(d) is plotted for
$q=0.75\Delta_0/v_F$.
Peaks at low frequency exist due to intraband electron transitions in the
conduction band. They start from
$\omega=0$
and disappear beyond
$\omega_1^{\text{sdw}}$
[the latter frequency is defined in
Eq.~(\ref{Eq::omega_1_sdw})].
Peaks close to
$\omega_2^{\text{sdw}}$
arise due to electron transitions from the valence band to the conduction
band shown in the inset in
Fig.~\ref{Fig::bands&peaks}c.
The frequency
$\omega_2^{\text{sdw}}$
is the threshold frequency for this processes. This energy is defined in
Eq.~(\ref{Eq::omega_2_sdw}).
	\label{Fig::SDW_sus}
	}
\end{figure*}

\subsection{Evaluation of the susceptibility tensor}

The above analysis demonstrates that, to characterize the neutron
scattering by electronic subsystem in the SDW state, one needs to know
$\tilde{\chi}^{\rm s,f}_{\bot}(\mathbf{Q},\omega)$
and
$\tilde{\chi}^{\rm s,f}_{xx}(\mathbf{Q},\omega)$.
We determine these quantities numerically using
Eq.~(\ref{Eq::kubo_four_Q_tilde}).
To be specific, our calculation in the SDW and SVHM phases are performed at
\begin{eqnarray}
	\label{Eq::doping_numerical}
	x=0.75N_F\Delta_0.
\end{eqnarray}
For such a doping level, parameters $\Delta$ and
$\Delta_{\sigma}$
deviate significantly from
$\Delta_0$.
Specifically, in the SDW phase, the order parameter and the chemical
potential
are~\cite{ourPRB_phasepAFM2017}
\begin{eqnarray}
\Delta\approx0.79\Delta_0,
\qquad
\mu\approx0.81\Delta_0.
\end{eqnarray}
We plot
$\tilde{\chi}^{\rm s,f}(\mathbf{Q},\omega)$
as a functions of the energy $\omega$ at the fixed wave vector
$\mathbf{Q}$,
see
Fig.~\ref{Fig::SDW_sus}.
Features that are of interest to us are most discernible when
$q\lesssim\Delta_0/v_F$.
We present
$\tilde{\chi}_{\bot}(\mathbf{Q},\omega)$
and
$\tilde{\chi}_{xx}(\mathbf{Q},\omega)$
for
$q=0.1\Delta_0/v_F$
in Fig.~\ref{Fig::SDW_sus}a,b and for
$q=0.75\Delta_0/v_F$
in Fig.~\ref{Fig::SDW_sus}c,d respectively.

Each component of the susceptibility tensor in
Fig.~\ref{Fig::SDW_sus}
has a peak at low frequency. It corresponds to electron-hole pairs in the
conduction band which are exited by neutrons. These peaks are localized
between
$\omega=0$
and a threshold frequency
$\omega_1^{\text{sdw}}$.
To derive
$\omega_1^{\text{sdw}}$
we write the energy conservation law for the intraband transitions
\begin{eqnarray}
	\label{Eq::intraband_energy_conservation}
	E^{(2)}_{\mathbf{k}+\mathbf{q}}-E^{(2)}_{\mathbf{k}}=\omega.
\end{eqnarray}
This must be solved together with
\begin{eqnarray}
	\label{Eq::region_k_lf_sdw}
	\theta(\mu-E^{(2)}_{\mathbf{k}})\theta(E^{(2)}_{\mathbf{k}+\mathbf{q}}-\mu)=1,
\end{eqnarray}
which is consequence of Fermi-Dirac statistics. In
Eq.~(\ref{Eq::region_k_lf_sdw})
the Heaviside step function is denoted as
$\theta(x)$.
System of
equations~(\ref{Eq::intraband_energy_conservation}),
(\ref{Eq::region_k_lf_sdw})
for the unknown variable
$\mathbf{k}$
has solutions only when
$0<\omega<\omega_1^{\text{sdw}}$,
where
\begin{eqnarray}
	\label{Eq::omega_1_sdw}
\omega_1^{\text{sdw}}=\sqrt{(\sqrt{\mu^2-\Delta^2}+q)^2+\Delta^2}-\mu.
\end{eqnarray}
Substituting specific values of $q$, one finds that
$\omega_1^{\text{sdw}} \approx 0.4\Delta_0$
when
$q=0.75\Delta_0/v_F$,
and
$\omega_1^{\text{sdw}} \approx 0.03\Delta_0$
when
$q=0.1\Delta_0/v_F$.
Both values for
$\omega_1^{\text{sdw}}$
are perfectly consistent with the numerical curves shown in
Fig.~\ref{Fig::SDW_sus}.

Another characteristic frequency
$\omega_2^{\text{sdw}}$
in
Fig.~\ref{Fig::SDW_sus}
is the threshold energy for the inelastic interband electron scattering,
see inset in
Fig.~\ref{Fig::bands&peaks}c.
When $\omega$ exceeds
$\omega_2^{\text{sdw}}$,
a new scattering channel opens, and the susceptibility becomes finite. This
is clearly visible on all panels of
Fig.~\ref{Fig::SDW_sus}.

To evaluate
$\omega_2^{\text{sdw}}$,
similar to our derivation of
Eq.~(\ref{Eq::omega_1_sdw}),
we use energy and momentum
conservation laws and Fermi-Dirac statistics of electrons to obtain
\begin{eqnarray}
	\label{Eq::omega_2_sdw}
	\omega_2^{\text{sdw}}=\Delta+\mu\approx1.6\Delta_0.
\end{eqnarray}
This equation has simple interpretation: it is exactly the energy necessary
to promote a quasiparticle from a state at the maximum of the completely
filled valence band to an empty state at the chemical potential level in
the partially filled conductance band. For chosen values of $\Delta$,
$\mu$, and $q$, such a transition is indeed consistent with both energy and
momentum conservation laws.

One can notice that for
$\omega > \omega_2^{\text{sdw}}$
the susceptibility tensor components pass through maximum near
$\omega_2^{\text{sdw}}$.
Locations of these peaks are slightly shifted to higher frequencies with
respect to
$\omega_2^{\text{sdw}}$.
This `blue shift', as well as a nonmonotonic behavior of susceptibility,
occurs due to structure of the joint density of states 
\begin{eqnarray}
\rho^{ss'}_{\sigma \sigma'} (\mathbf{q},\omega)
=
\int \delta(
	E_{\sigma\mathbf{k}}^{(s)}
	-
	E_{\sigma'\mathbf{k}+\mathbf{q}}^{(s')}
	-\omega) 
\frac{d^3\mathbf{k}}{(2\pi)^3}.
	\label{Eq::jdos}
\end{eqnarray}
We see in
Fig.~\ref{Fig::SDW_sus}d
that
$\tilde{\chi}^{\rm s}_{xx}({\bf Q}, \omega)$,
as a function of
$(\omega - \omega_2^{\text{sdw}})$,
demonstrates slower (linear) growth than
$\tilde{\chi}^{\rm f}_{xx}({\bf Q}, \omega)$.
This is a consequence of nullification of the matrix element for the
corresponding interband electron transition exactly at
$\omega = \omega_2^{\text{sdw}}$.
The matrix element becomes non-zero when
$\omega>\omega_2^{\text{sdw}}$,
however, its value remains small for small $q$. Thus,
$\tilde{\chi}^{\rm s}_{xx}({\bf Q}, \omega)$
in
Fig.~\ref{Fig::SDW_sus}b
is very close to zero in contrast with
$\tilde{\chi}^{\rm s}_{xx}({\bf Q}, \omega)$
in
Fig.~\ref{Fig::SDW_sus}d.
\section{spin susceptibility of the spin-valley half-metallic phase}
\label{sec::svhm}
According to
Eq.~(\ref{Eq::av_spin_svhm}),
in the SVHM state both
$\langle S_x(\mathbf{r})\rangle$
and
$\langle S_y(\mathbf{r})\rangle$
are non-zero. Therefore, the rotational symmetry around $x$-axis is broken,
and all diagonal components of the susceptibility tensor may differ from
each other. As for off-diagonal components, they remain zero. This is a
consequence of the electron-hole symmetry of the
bands~(\ref{Eq::disperse_no_order}).
If the electron and hole valleys are asymmetrical, the off-diagonal
elements acquire finite values. Specifically, we calculate separately all
diagonal components of the susceptibility tensor, assuming
$m_a=m_b$
in band
structure~(\ref{Eq::disperse_no_order}).
Then we introduce an asymmetry between the electron and hole
bands~(\ref{Eq::disperse_no_order})
through difference in effective masses
$m_a\neq m_b$
and estimate components
$\tilde{\chi}_{xy}(\mathbf{Q},\omega)$
and
$\tilde{\chi}_{yx}(\mathbf{Q},\omega)$.

For numerical calculations we take the doping value, defined in
Eq.~(\ref{Eq::doping_numerical}).
We obtain the order parameter
$\Delta_{\sigma}=0.5\Delta_0$
from
Eq.~(\ref{Eq::order_par_svhm})
and the chemical potential
$\mu=0.625\Delta_0$
using Eq.~(11) of
Ref.~\onlinecite{ourPRL_half-met2017}.
\begin{figure*}
	\center{\includegraphics[width=1\linewidth]{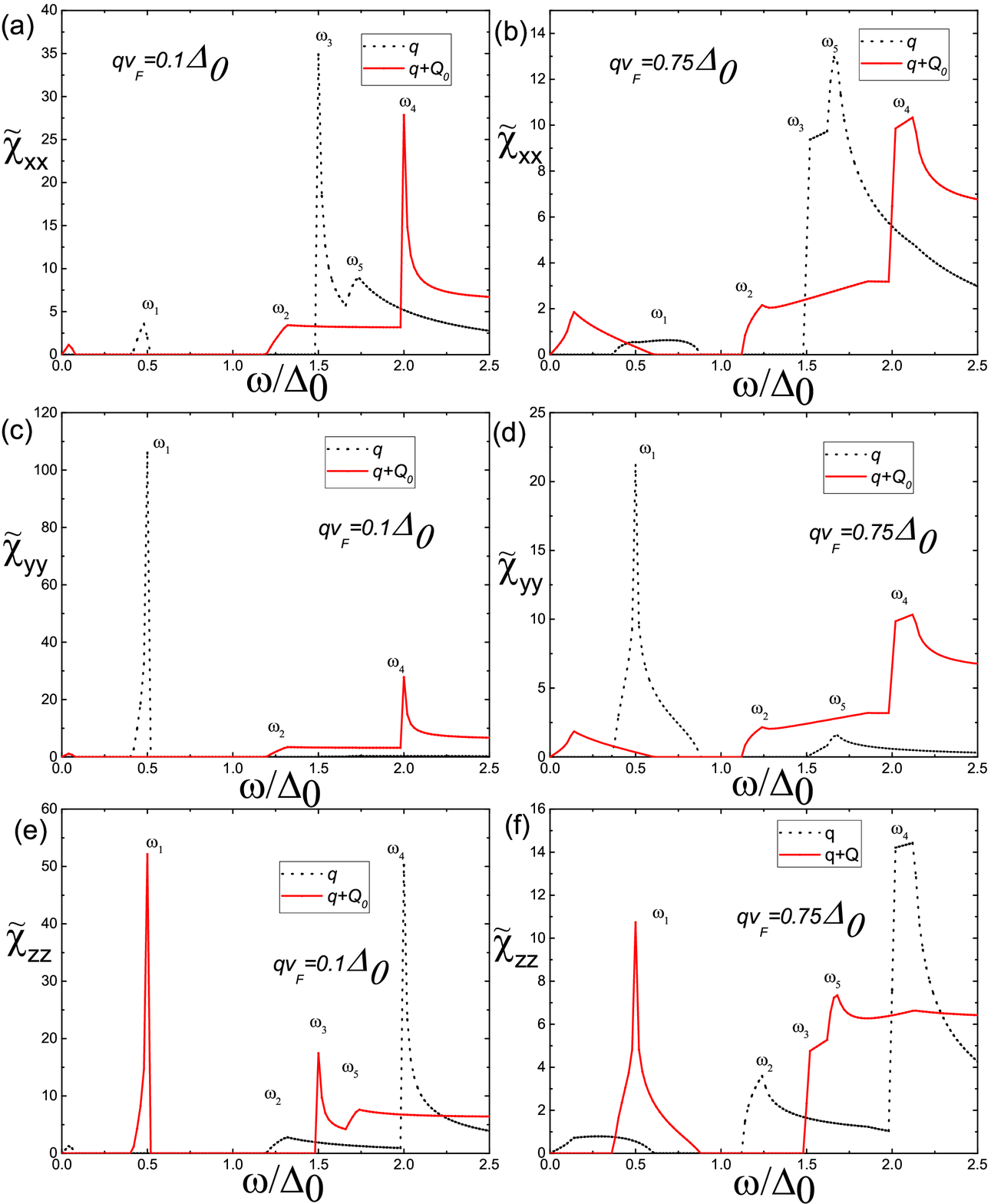}}
	\caption{The diagonal components of the susceptibility tensor
$\tilde{\chi}^{\rm s,f}(\mathbf{Q},\omega)$
at fixed momentum
$\mathbf{Q}$
versus frequency
$\omega$
in the SVHM phase. Dot curves show slow part of the susceptibility
calculated for
$\mathbf{Q}=\mathbf{q}$;
solid curves show fast part of the susceptibility calculated for
$\mathbf{Q}=\mathbf{q}\pm\mathbf{Q}_0$.
Panels (a) and~(b) present
$\tilde{\chi}_{xx}(\mathbf{Q},\omega)$,
panels~(c) and~(d) present
$\tilde{\chi}_{yy}(\mathbf{Q},\omega)$,
and panels~(e) and~(f) present
$\tilde{\chi}_{zz}(\mathbf{Q},\omega)$
correspondingly. The data in panels~(a),~(c),
and~(e)
is plotted for
$q=0.1\Delta_0/v_F$.
The data in panels~(b),
(d) and~(f) is plotted for
$q=0.75\Delta_0/v_F$.
The (weak) peaks at the lowest frequency are due to intraband transitions
within the conduction band. All other peaks are caused by transitions
between bands. These transitions are marked by arrows in the inset of
Fig.~\ref{Fig::bands&peaks}d.
Each threshold frequency
$\omega_{1}$,
$\ldots$,
$\omega_{5}$
represents the opening of a new interband scattering channel. They are
determined by
Eq.~(\ref{Eq::svhm_omega_n}).
	\label{Fig::SVHM_sus_diag}
	}
\end{figure*}

\subsection{Diagonal components of the susceptibility tensor}
\label{subsec::diag_components}

The dependence of the diagonal components on frequency $\omega$ is
presented in
Fig.~\ref{Fig::SVHM_sus_diag}.
As we already mentioned in
Sec.~\ref{subsec::gen_remarks},
within our model, the tensor components are insensitive to the direction of
${\bf q}$,
only the absolute value of the transferred momentum $q$ matters. To
illustrate the dependence on $q$, the curves in
panels~\ref{Fig::SVHM_sus_diag}(a), (c), and (e)
are plotted for
$q=0.1\Delta_0/v_F$,
other three panels show the diagonal components at
$q=0.75\Delta_0/v_F$

The peaks that start from zero frequency are due to the intraband
scattering processes. They are similar to the low-energy peaks discussed in
Sec.~\ref{sec::sdw}
in the context of the SDW.

In
Fig.~\ref{Fig::SVHM_sus_diag},
finite-frequency spectral features (peaks or steps), marked by
$\omega_1, \ldots, \omega_5$,
arise due to the interband electron transitions. These transitions are
illustrated in the inset in
Fig.~\ref{Fig::bands&peaks}(d).
Frequencies
$\omega_{n}$
may be found in the same manner as
$\omega_2^{\text{sdw}}$.
One needs to find minimum frequency at which a solution of the equation
\begin{eqnarray}
	\label{Eq::omega_n}
	\omega_{n}=E_f-E_i
\end{eqnarray}
still exists. In
Eq.~(\ref{Eq::omega_n})
$E_f$
and
$E_i$
are final and initial energies of the excited electron. State at
$E_f$
must be empty and state at
$E_i$
must be occupied in the ground state to allow excitation process. For five
excitations channels shown in
Fig.~\ref{Fig::SVHM_sus_diag},
we derive five
threshold frequencies. For doping level given by
Eq.~(\ref{Eq::doping_numerical})
these frequencies are
\begin{eqnarray}
\label{Eq::svhm_omega_n}
\omega_1
&=&
\sqrt{\max(\sqrt{\mu^2-\Delta_{\sigma}^2}-v_Fq,0)^2+\Delta_0^2} -\mu,
\nonumber\\ 
\omega_2
&=&
\mu+\sqrt{\max(\sqrt{\mu^2-\Delta_{\sigma}^2}-v_Fq,0)^2+\Delta_{\sigma}^2},
\nonumber\\
\omega_3 &=& \Delta_0+\Delta_{\sigma},
\nonumber\\ 
\omega_4&=&2\Delta_0,
\nonumber\\
\omega_5&=&\mu+\sqrt{\max(\sqrt{\mu^2-\Delta_{\sigma}^2}-v_Fq,0)^2+\Delta_0^2}. 
\end{eqnarray}
Numerical value of the above frequencies is in the perfect agreement with the threshold frequencies in Fig.~\ref{Fig::SVHM_sus_diag}.
As a result of a more complex band structure, the SVHM neutron scattering
spectrum has richer structure than the spectrum of the SDW. In
Fig.~\ref{Fig::SVHM_sus_diag}(a), (c), and (e),
which represent the spectra for
$q= 0.1 \Delta_0 /v_{\rm F}$,
one can discern three spectral peaks (at
$\omega_1$,
$\omega_3$,
and
$\omega_4$),
and two step-like features (at
$\omega_2$,
and
$\omega_5$).
The intensities of these spectral components demonstrate non-trivial
dependence on spin polarization. At higher $q$, the peaks broaden and
merge. However, the characteristic frequencies
remain discernible
even in such a regime.

\subsection{Off-diagonal components of the susceptibility tensor}

Besides the structure of the diagonal components of the susceptibility
tensor, the SVHM has yet another distinction that separates it from the
SDW. As we already pointed out above, the SVHM phase may have finite values
of the off-diagonal components
$\tilde{\chi}_{xy}$
and
$\tilde{\chi}_{yx}$
when the perfect electron-hole symmetry is broken. (For the SDW, these
components vanish due to the rotation symmetry.)

\begin{figure}[t]
\center{\includegraphics[width=1\linewidth]{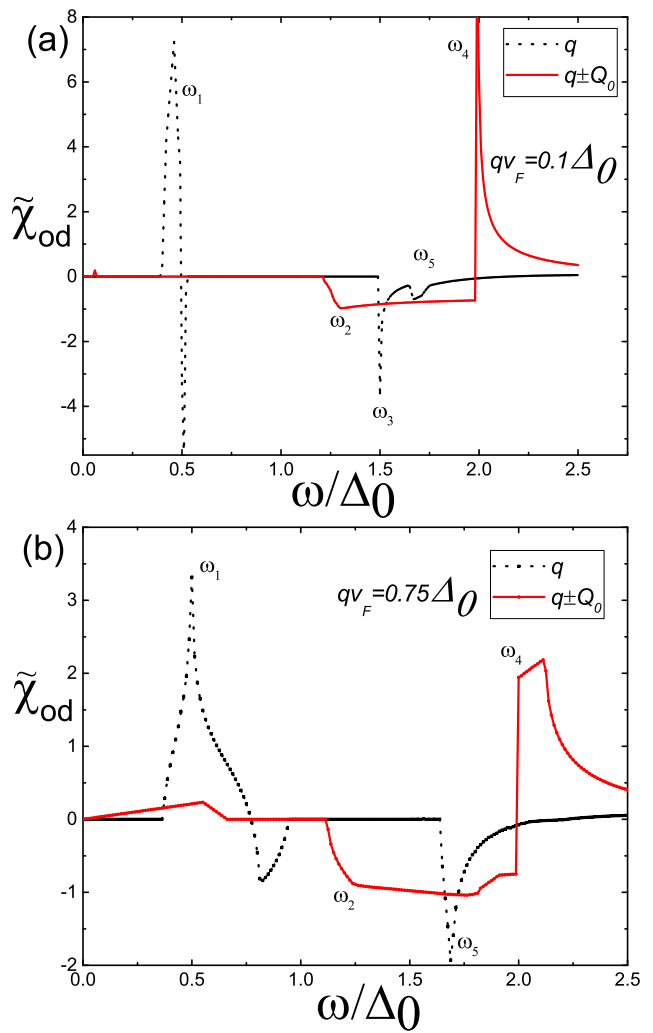}}
\caption{The off-diagonal component of the susceptibility tensor
$\tilde{\chi}^{\rm s,f}_{xy}(\mathbf{Q},\omega)$
at fixed momentum
$\mathbf{Q}$
versus frequency $\omega$ in the SVHM phase. Dot curves show the slow part
of the susceptibility calculated for
$\mathbf{Q}=\mathbf{q}$;
solid curves show the fast part of the susceptibility calculated for
$\mathbf{Q}=\mathbf{q}\pm\mathbf{Q}_0$.
The data in panel~(a) is plotted for
$q=0.1\Delta_0/v_F$.
The data in panel~(b) is plotted for
$q=0.75\Delta_0/v_F$.
Each threshold frequency
$\omega_{1},\ldots, \omega_{5}$
represents the opening of a new interband scattering channel. They are
determined by
Eq.~(\ref{Eq::svhm_omega_n}).
\label{Fig::SVHM_sus_offdiag}
}
\end{figure}

To study the off-diagonal components, we break the electron-hole symmetry
in our model by introducing the difference between the effective masses of
electrons and holes
\begin{eqnarray}
\frac{m_b-m_a}{m_b+m_a}=0.1.
\end{eqnarray}
In such a regime, we evaluate the off-diagonal components numerically.

Figure~\ref{Fig::SVHM_sus_offdiag} shows the dependence
$\tilde{\chi}_{xy}$
on frequency $\omega$.
Panel~\ref{Fig::SVHM_sus_offdiag}(a) is plotted for
$q=0.1\Delta_0/v_F$,
and panel~\ref{Fig::SVHM_sus_offdiag}(b)
is plotted for
$q=0.75\Delta_0/v_F$.
As for
$\tilde{\chi}_{yx}$,
it can be determined using the relation:
\begin{eqnarray}
	\label{Eq::off_diag_relation}
	\tilde{\chi}_{xy}(\mathbf{Q},\omega)=-\tilde{\chi}_{yx}(\mathbf{Q},\omega).
\end{eqnarray}
This equality can be derived by substitution of
Eq.~(\ref{Eq::spin_f_and_s})
into
Eq.~(\ref{Eq::kubo_four_Q_tilde}),
and it follows from the anticommutation rule for Pauli matrices
$\sigma_x\sigma_y=-\sigma_y\sigma_x$.
Note also that, unlike
$\tilde \chi_{\alpha \alpha}^{\rm s,f}$,
which cannot be negative, the off-diagonal components of the tensor are not
constrained by such a requirement, and
$\tilde{\chi}_{xy}^{\rm s,f}$
can be of either sign, as indeed seen in
Fig.~\ref{Fig::SVHM_sus_offdiag}.

The same five scattering channels depicted in the inset of
Fig.~\ref{Fig::bands&peaks}(d)
control the structure of the off-diagonal components. We see peaks at
frequencies
$\omega_{1}$,
$\omega_{3}$,
$\omega_{4}$,
and
$\omega_{5}$,
as well as a step-like feature at
$\omega_{2}$.
The intensities of the peaks at
$\omega_3$
and
$\omega_5$
are particularly sensitive to $q$ in the chosen range of parameters. One
can also notice that the peak at
$\omega_{1}$
acquires `a shadow' peak with the opposite sign at somewhat higher
$\omega$,

\section{Discussion and conclusions}
\label{sec::discussion}

Our study is motivated by the question if the neutron scattering can be
used to distinguish the SDW and the SVHM phases. With this aim in mind, we
calculated the dynamical spin susceptibility tensor of both phases as a
function of frequency at fixed momentum. It is demonstrated that the
susceptibilities of the SDW and SVHM have two well-noticeable qualitative
distinctions.

The first appreciable difference is a number of high peaks in the diagonal
components of the tensor. The SDW phase has only one large peak. At the
same time the SVHM has three high and two weaker and broader features.
Since each peak represents an interband transition of electrons, the
distinction in the number of peaks can be traced to the number of
quasiparticle bands in these phases: four bands in the SVHM phase versus
only two bands in the SDW phase.

The second difference occurs if electron-hole symmetry is broken (we model
such an asymmetry by introducing non-identical effective masses of electrons
and holes). In this case, the SVHM has finite off-diagonal components of
the susceptibility tensor. In the SDW phase these components remain zero
(in the SDW phase this property is robust since it is protected by the
rotation symmetry around order parameter polarization axis).

The intensity of a specific peak depends on the joint density of states and
on a corresponding matrix element. Obviously, both these quantities are
functions of $\omega$ and $q$.
In addition, matrix elements
depend also on
axis labels [for example, the matrix element for
$\chi_{yy}$
is not necessary equal to the matrix element for
$\chi_{zz}$].
Consequently, the intensity of a peak representing a specific inelastic
scattering channel is sensitive to polarization. For example, the peak at
$\omega=\omega_1$
in
$\chi^{\rm s}_{xx}$
is much weaker than the same peak in
$\chi^{\rm s}_{yy}$,
see
Fig.~\ref{Fig::SVHM_sus_diag}a,c.
As for
$\chi^{\rm s}_{zz}$,
it demonstrates no peak at
$\omega=\omega_1$,
see
Fig.~\ref{Fig::SVHM_sus_diag}(e).

In our model, the joint density of states (\ref{Eq::jdos}) diverges when the Pauli principle
allows for transition between the edges of the bands. This is the case for
the transitions at
$\omega=\omega_{1,3,4}$.
If large joint density of states is accompanied by a finite matrix element,
the peak intensity is particularly strong. For example, matrix elements at
$\omega_1$ and $\omega_4$ in
Fig.~\ref{Fig::SVHM_sus_diag}(e) are large, while matrix element at
$\omega_3$ in the same panel is low.

Our calculations were performed for commensurate homogeneous phases only.
However, we expect that qualitative behavior of the dynamical spin
susceptibility survives in the incommensurate phases as well. Indeed, the
most pronounced qualitative differences between the spectra of the
commensurate phases are associated either with
equality~\eqref{Eq::sdw_sus_relation_2}
or with the fact that the SDW state has two single-electron bands while the
SVHM state hosts four such bands. Both these properties survive in the
incommensurate phases. Specifically, the
relation~\eqref{Eq::sdw_sus_relation_2}
is a consequence of the uniaxial spin-rotation invariance of the SDW phase,
which holds regardless of the commensurability of the order parameter. As
for the band structure of the incommensurate SVHM phase, the expressions
for its four bands were derived in
Ref.~\onlinecite{ourPRB_half-met2018},
see Eq.~(56) there. Thus, we expect that the inelastic neutron scattering
can be used to detect the SVHM state even for the incommensurate order
parameter.

In conclusion, we calculated the dynamical spin susceptibility for the
doped spin-density wave state and the doped spin-valley half-metallic
state at different momenta. Due to more complex band structure, the
SVHM spin susceptibility tensor demonstrates richer frequency dependence,
and may have finite off-diagonal components. Our analysis shows that
the inelastic neutron scattering may be used to distinguish the SDW and the
SVHM phases in experiment.

\section{acknowledgements}
This work was partially supported by the JSPS-Russian Foundation for Basic
Research joint Project
No.~19-52-50015.
One of us (DAH) was partially supported by the Foundation for the
Advancement of Theoretical Physics and Mathematics ``BASIS".

\begin{thebibliography}{40}
\expandafter\ifx\csname natexlab\endcsname\relax\def\natexlab#1{#1}\fi
\expandafter\ifx\csname bibnamefont\endcsname\relax
  \def\bibnamefont#1{#1}\fi
\expandafter\ifx\csname bibfnamefont\endcsname\relax
  \def\bibfnamefont#1{#1}\fi
\expandafter\ifx\csname citenamefont\endcsname\relax
  \def\citenamefont#1{#1}\fi

\bibitem[{\citenamefont{Rice}(1970)}]{rice1970}
\bibinfo{author}{\bibfnamefont{T.~M.} \bibnamefont{Rice}},
  {``}\bibinfo{title}{Band-Structure Effects in Itinerant
  Antiferromagnetism},{''} \bibinfo{journal}{Phys. Rev. B}
  \textbf{\bibinfo{volume}{2}}, \bibinfo{pages}{3619} (\bibinfo{year}{1970}).

\bibitem[{\citenamefont{Rakhmanov et~al.}(2013)\citenamefont{Rakhmanov,
  Rozhkov, Sboychakov, and Nori}}]{ourPRB_phasepAFM2017}
\bibinfo{author}{\bibfnamefont{A.~L.} \bibnamefont{Rakhmanov}},
  \bibinfo{author}{\bibfnamefont{A.~V.} \bibnamefont{Rozhkov}},
  \bibinfo{author}{\bibfnamefont{A.~O.} \bibnamefont{Sboychakov}},
  \bibnamefont{and} \bibinfo{author}{\bibfnamefont{F.}~\bibnamefont{Nori}},
  {``}\bibinfo{title}{Phase separation of antiferromagnetic ground states in
  systems with imperfect nesting},{''} \bibinfo{journal}{Phys. Rev. B}
  \textbf{\bibinfo{volume}{87}}, \bibinfo{pages}{075128}
  (\bibinfo{year}{2013}).

\bibitem[{\citenamefont{Rozhkov et~al.}(2017)\citenamefont{Rozhkov, Rakhmanov,
  Sboychakov, Kugel, and Nori}}]{ourPRL_half-met2017}
\bibinfo{author}{\bibfnamefont{A.~V.} \bibnamefont{Rozhkov}},
  \bibinfo{author}{\bibfnamefont{A.~L.} \bibnamefont{Rakhmanov}},
  \bibinfo{author}{\bibfnamefont{A.~O.} \bibnamefont{Sboychakov}},
  \bibinfo{author}{\bibfnamefont{K.~I.} \bibnamefont{Kugel}}, \bibnamefont{and}
  \bibinfo{author}{\bibfnamefont{F.}~\bibnamefont{Nori}},
  {``}\bibinfo{title}{Spin-Valley Half-Metal as a Prospective Material for Spin
  Valleytronics},{''} \bibinfo{journal}{Phys. Rev. Lett.}
  \textbf{\bibinfo{volume}{119}}, \bibinfo{pages}{107601}
  (\bibinfo{year}{2017}).

\bibitem[{\citenamefont{Chuang et~al.}(2001)\citenamefont{Chuang, Gromko,
  Dessau, Kimura, and Tokura}}]{Chuang1509}
\bibinfo{author}{\bibfnamefont{Y.-D.} \bibnamefont{Chuang}},
  \bibinfo{author}{\bibfnamefont{A.~D.} \bibnamefont{Gromko}},
  \bibinfo{author}{\bibfnamefont{D.~S.} \bibnamefont{Dessau}},
  \bibinfo{author}{\bibfnamefont{T.}~\bibnamefont{Kimura}}, \bibnamefont{and}
  \bibinfo{author}{\bibfnamefont{Y.}~\bibnamefont{Tokura}},
  {``}\bibinfo{title}{Fermi Surface Nesting and Nanoscale Fluctuating
  Charge/Orbital Ordering in Colossal Magnetoresistive Oxides},{''}
  \bibinfo{journal}{Science} \textbf{\bibinfo{volume}{292}},
  \bibinfo{pages}{1509} (\bibinfo{year}{2001}).

\bibitem[{\citenamefont{Kiesel et~al.}(2012)\citenamefont{Kiesel, Platt, Hanke,
  Abanin, and Thomale}}]{PhysRevB.86.020507}
\bibinfo{author}{\bibfnamefont{M.~L.} \bibnamefont{Kiesel}},
  \bibinfo{author}{\bibfnamefont{C.}~\bibnamefont{Platt}},
  \bibinfo{author}{\bibfnamefont{W.}~\bibnamefont{Hanke}},
  \bibinfo{author}{\bibfnamefont{D.~A.} \bibnamefont{Abanin}},
  \bibnamefont{and} \bibinfo{author}{\bibfnamefont{R.}~\bibnamefont{Thomale}},
  {``}\bibinfo{title}{Competing many-body instabilities and unconventional
  superconductivity in graphene},{''} \bibinfo{journal}{Phys. Rev. B}
  \textbf{\bibinfo{volume}{86}}, \bibinfo{pages}{020507}
  (\bibinfo{year}{2012}).

\bibitem[{\citenamefont{Gor'kov and Teitel'baum}(2010)}]{PhysRevB.82.020510}
\bibinfo{author}{\bibfnamefont{L.~P.} \bibnamefont{Gor'kov}} \bibnamefont{and}
  \bibinfo{author}{\bibfnamefont{G.~B.} \bibnamefont{Teitel'baum}},
  {``}\bibinfo{title}{Spatial inhomogeneities in iron pnictide superconductors:
  The formation of charge stripes},{''} \bibinfo{journal}{Phys. Rev. B}
  \textbf{\bibinfo{volume}{82}}, \bibinfo{pages}{020510}
  (\bibinfo{year}{2010}).

\bibitem[{\citenamefont{\ifmmode~\check{S}\else \v{S}\fi{}imkovic
  et~al.}(2016)\citenamefont{\ifmmode~\check{S}\else \v{S}\fi{}imkovic, Liu,
  Deng, and Kozik}}]{PhysRevB.94.085106}
\bibinfo{author}{\bibfnamefont{F.}~\bibnamefont{\ifmmode~\check{S}\else
  \v{S}\fi{}imkovic}}, \bibinfo{author}{\bibfnamefont{X.-W.}
  \bibnamefont{Liu}}, \bibinfo{author}{\bibfnamefont{Y.}~\bibnamefont{Deng}},
  \bibnamefont{and} \bibinfo{author}{\bibfnamefont{E.}~\bibnamefont{Kozik}},
  {``}\bibinfo{title}{Ground-state phase diagram of the repulsive fermionic
  $t\ensuremath{-}{t}^{\ensuremath{'}}$ Hubbard model on the square lattice
  from weak coupling},{''} \bibinfo{journal}{Phys. Rev. B}
  \textbf{\bibinfo{volume}{94}}, \bibinfo{pages}{085106}
  (\bibinfo{year}{2016}).

\bibitem[{\citenamefont{Mosoyan et~al.}(2018)\citenamefont{Mosoyan, Rozhkov,
  Sboychakov, and Rakhmanov}}]{mosoyan_aa_graphit2018}
\bibinfo{author}{\bibfnamefont{K.~S.} \bibnamefont{Mosoyan}},
  \bibinfo{author}{\bibfnamefont{A.~V.} \bibnamefont{Rozhkov}},
  \bibinfo{author}{\bibfnamefont{A.~O.} \bibnamefont{Sboychakov}},
  \bibnamefont{and} \bibinfo{author}{\bibfnamefont{A.~L.}
  \bibnamefont{Rakhmanov}}, {``}\bibinfo{title}{Spin-density wave state in
  simple hexagonal graphite},{''} \bibinfo{journal}{Phys. Rev. B}
  \textbf{\bibinfo{volume}{97}}, \bibinfo{pages}{075131}
  (\bibinfo{year}{2018}).

\bibitem[{\citenamefont{Rakhmanov et~al.}(2018)\citenamefont{Rakhmanov,
  Sboychakov, Kugel, Rozhkov, and Nori}}]{ourPRB_half-met2018}
\bibinfo{author}{\bibfnamefont{A.~L.} \bibnamefont{Rakhmanov}},
  \bibinfo{author}{\bibfnamefont{A.~O.} \bibnamefont{Sboychakov}},
  \bibinfo{author}{\bibfnamefont{K.~I.} \bibnamefont{Kugel}},
  \bibinfo{author}{\bibfnamefont{A.~V.} \bibnamefont{Rozhkov}},
  \bibnamefont{and} \bibinfo{author}{\bibfnamefont{F.}~\bibnamefont{Nori}},
  {``}\bibinfo{title}{Spin-valley half-metal in systems with Fermi surface
  nesting},{''} \bibinfo{journal}{Phys. Rev. B} \textbf{\bibinfo{volume}{98}},
  \bibinfo{pages}{155141} (\bibinfo{year}{2018}).

\bibitem[{\citenamefont{Nandkishore
  et~al.}(2012{\natexlab{a}})\citenamefont{Nandkishore, Chern, and
  Chubukov}}]{Nandkishore2012}
\bibinfo{author}{\bibfnamefont{R.}~\bibnamefont{Nandkishore}},
  \bibinfo{author}{\bibfnamefont{G.-W.} \bibnamefont{Chern}}, \bibnamefont{and}
  \bibinfo{author}{\bibfnamefont{A.~V.} \bibnamefont{Chubukov}},
  {``}\bibinfo{title}{Itinerant Half-Metal Spin-Density-Wave State on the
  Hexagonal Lattice},{''} \bibinfo{journal}{Phys. Rev. Lett.}
  \textbf{\bibinfo{volume}{108}}, \bibinfo{pages}{227204}
  (\bibinfo{year}{2012}{\natexlab{a}}).

\bibitem[{\citenamefont{Sboychakov et~al.}(2017)\citenamefont{Sboychakov,
  Rakhmanov, Kugel, Rozhkov, and Nori}}]{our_PRB_magfield_imp_nest}
\bibinfo{author}{\bibfnamefont{A.~O.} \bibnamefont{Sboychakov}},
  \bibinfo{author}{\bibfnamefont{A.~L.} \bibnamefont{Rakhmanov}},
  \bibinfo{author}{\bibfnamefont{K.~I.} \bibnamefont{Kugel}},
  \bibinfo{author}{\bibfnamefont{A.~V.} \bibnamefont{Rozhkov}},
  \bibnamefont{and} \bibinfo{author}{\bibfnamefont{F.}~\bibnamefont{Nori}},
  {``}\bibinfo{title}{Magnetic field effects in electron systems with imperfect
  nesting},{''} \bibinfo{journal}{Phys. Rev. B} \textbf{\bibinfo{volume}{95}},
  \bibinfo{pages}{014203} (\bibinfo{year}{2017}).

\bibitem[{\citenamefont{Gonz\'alez and
  Stauber}(2019)}]{gonzalez_kohn-lutt_twist2019}
\bibinfo{author}{\bibfnamefont{J.}~\bibnamefont{Gonz\'alez}} \bibnamefont{and}
  \bibinfo{author}{\bibfnamefont{T.}~\bibnamefont{Stauber}},
  {``}\bibinfo{title}{Kohn-Luttinger Superconductivity in Twisted Bilayer
  Graphene},{''} \bibinfo{journal}{Phys. Rev. Lett.}
  \textbf{\bibinfo{volume}{122}}, \bibinfo{pages}{026801}
  (\bibinfo{year}{2019}).

\bibitem[{\citenamefont{Sboychakov et~al.}(2018)\citenamefont{Sboychakov,
  Rozhkov, Rakhmanov, and Nori}}]{twist_graph_bias_gap_prl2018}
\bibinfo{author}{\bibfnamefont{A.~O.} \bibnamefont{Sboychakov}},
  \bibinfo{author}{\bibfnamefont{A.~V.} \bibnamefont{Rozhkov}},
  \bibinfo{author}{\bibfnamefont{A.~L.} \bibnamefont{Rakhmanov}},
  \bibnamefont{and} \bibinfo{author}{\bibfnamefont{F.}~\bibnamefont{Nori}},
  {``}\bibinfo{title}{Externally Controlled Magnetism and Band Gap in Twisted
  Bilayer Graphene},{''} \bibinfo{journal}{Phys. Rev. Lett.}
  \textbf{\bibinfo{volume}{120}}, \bibinfo{pages}{266402}
  (\bibinfo{year}{2018}).

\bibitem[{\citenamefont{Rakhmanov et~al.}(2017)\citenamefont{Rakhmanov, Kugel,
  Kagan, Rozhkov, and Sboychakov}}]{nesting_review2017}
\bibinfo{author}{\bibfnamefont{A.~L.} \bibnamefont{Rakhmanov}},
  \bibinfo{author}{\bibfnamefont{K.~I.} \bibnamefont{Kugel}},
  \bibinfo{author}{\bibfnamefont{M.~Y.} \bibnamefont{Kagan}},
  \bibinfo{author}{\bibfnamefont{A.~V.} \bibnamefont{Rozhkov}},
  \bibnamefont{and} \bibinfo{author}{\bibfnamefont{A.~O.}
  \bibnamefont{Sboychakov}}, {``}\bibinfo{title}{Inhomogeneous electron states
  in the systems with imperfect nesting},{''} \bibinfo{journal}{JETP Lett.}
  \textbf{\bibinfo{volume}{105}}, \bibinfo{pages}{806} (\bibinfo{year}{2017}).

\bibitem[{\citenamefont{Akzyanov et~al.}(2014)\citenamefont{Akzyanov,
  Sboychakov, Rozhkov, Rakhmanov, and Nori}}]{aa_graph2014}
\bibinfo{author}{\bibfnamefont{R.~S.} \bibnamefont{Akzyanov}},
  \bibinfo{author}{\bibfnamefont{A.~O.} \bibnamefont{Sboychakov}},
  \bibinfo{author}{\bibfnamefont{A.~V.} \bibnamefont{Rozhkov}},
  \bibinfo{author}{\bibfnamefont{A.~L.} \bibnamefont{Rakhmanov}},
  \bibnamefont{and} \bibinfo{author}{\bibfnamefont{F.}~\bibnamefont{Nori}},
  {``}\bibinfo{title}{{AA}-stacked bilayer graphene in an applied electric
  field: Tunable antiferromagnetism and coexisting exciton order
  parameter},{''} \bibinfo{journal}{Phys. Rev. B}
  \textbf{\bibinfo{volume}{90}}, \bibinfo{pages}{155415}
  (\bibinfo{year}{2014}).

\bibitem[{\citenamefont{Sboychakov
  et~al.}(2013{\natexlab{a}})\citenamefont{Sboychakov, Rakhmanov, Rozhkov, and
  Nori}}]{aa_graph2013}
\bibinfo{author}{\bibfnamefont{A.~O.} \bibnamefont{Sboychakov}},
  \bibinfo{author}{\bibfnamefont{A.~L.} \bibnamefont{Rakhmanov}},
  \bibinfo{author}{\bibfnamefont{A.~V.} \bibnamefont{Rozhkov}},
  \bibnamefont{and} \bibinfo{author}{\bibfnamefont{F.}~\bibnamefont{Nori}},
  {``}\bibinfo{title}{Metal-insulator transition and phase separation in doped
  {AA}-stacked graphene bilayer},{''} \bibinfo{journal}{Phys. Rev. B}
  \textbf{\bibinfo{volume}{87}}, \bibinfo{pages}{121401}
  (\bibinfo{year}{2013}{\natexlab{a}}).

\bibitem[{\citenamefont{Sboychakov
  et~al.}(2013{\natexlab{b}})\citenamefont{Sboychakov, Rozhkov, Kugel,
  Rakhmanov, and Nori}}]{phasep_pnics2013}
\bibinfo{author}{\bibfnamefont{A.~O.} \bibnamefont{Sboychakov}},
  \bibinfo{author}{\bibfnamefont{A.~V.} \bibnamefont{Rozhkov}},
  \bibinfo{author}{\bibfnamefont{K.~I.} \bibnamefont{Kugel}},
  \bibinfo{author}{\bibfnamefont{A.~L.} \bibnamefont{Rakhmanov}},
  \bibnamefont{and} \bibinfo{author}{\bibfnamefont{F.}~\bibnamefont{Nori}},
  {``}\bibinfo{title}{Electronic phase separation in iron pnictides},{''}
  \bibinfo{journal}{Phys. Rev. B} \textbf{\bibinfo{volume}{88}},
  \bibinfo{pages}{195142} (\bibinfo{year}{2013}{\natexlab{b}}).

\bibitem[{\citenamefont{Rozhkov}(2009{\natexlab{a}})}]{q1d2009}
\bibinfo{author}{\bibfnamefont{A.~V.} \bibnamefont{Rozhkov}},
  {``}\bibinfo{title}{Superconductivity without attraction in a
  quasi-one-dimensional metal},{''} \bibinfo{journal}{Phys. Rev. B}
  \textbf{\bibinfo{volume}{79}}, \bibinfo{pages}{224520}
  (\bibinfo{year}{2009}{\natexlab{a}}).

\bibitem[{\citenamefont{Rozhkov}(2009{\natexlab{b}})}]{q1d2009_2}
\bibinfo{author}{\bibfnamefont{A.~V.} \bibnamefont{Rozhkov}},
  {``}\bibinfo{title}{Competition between different order parameters in a
  quasi-one-dimensional superconductor},{''} \bibinfo{journal}{Phys. Rev. B}
  \textbf{\bibinfo{volume}{79}}, \bibinfo{pages}{224501}
  (\bibinfo{year}{2009}{\natexlab{b}}).

\bibitem[{\citenamefont{Rozhkov}(2003)}]{q1d2003}
\bibinfo{author}{\bibfnamefont{A.~V.} \bibnamefont{Rozhkov}},
  {``}\bibinfo{title}{Variational description of the dimensional crossover in
  an array of coupled one-dimensional conductors},{''} \bibinfo{journal}{Phys.
  Rev. B} \textbf{\bibinfo{volume}{68}}, \bibinfo{pages}{115108}
  (\bibinfo{year}{2003}).

\bibitem[{\citenamefont{Hirschfeld et~al.}(2011)\citenamefont{Hirschfeld,
  Korshunov, and Mazin}}]{hirschfeld_review2011}
\bibinfo{author}{\bibfnamefont{P.}~\bibnamefont{Hirschfeld}},
  \bibinfo{author}{\bibfnamefont{M.}~\bibnamefont{Korshunov}},
  \bibnamefont{and} \bibinfo{author}{\bibfnamefont{I.}~\bibnamefont{Mazin}},
  {``}\bibinfo{title}{Gap symmetry and structure of Fe-based
  superconductors},{''} \bibinfo{journal}{Rep. Prog. Phys.}
  \textbf{\bibinfo{volume}{74}}, \bibinfo{pages}{124508}
  (\bibinfo{year}{2011}).

\bibitem[{\citenamefont{Fernandes and
  Schmalian}(2010)}]{fernandez_pnic_so5_2010}
\bibinfo{author}{\bibfnamefont{R.~M.} \bibnamefont{Fernandes}}
  \bibnamefont{and}
  \bibinfo{author}{\bibfnamefont{J.}~\bibnamefont{Schmalian}},
  {``}\bibinfo{title}{Competing order and nature of the pairing state in the
  iron pnictides},{''} \bibinfo{journal}{Phys. Rev. B}
  \textbf{\bibinfo{volume}{82}}, \bibinfo{pages}{014521}
  (\bibinfo{year}{2010}).

\bibitem[{\citenamefont{Gr{\"u}ner}(1994)}]{gruner_book}
\bibinfo{author}{\bibfnamefont{G.}~\bibnamefont{Gr{\"u}ner}},
  \emph{\bibinfo{title}{Density Waves In Solids}}
  (\bibinfo{publisher}{Addison-Wesley Publishing Company},
  \bibinfo{year}{1994}).

\bibitem[{\citenamefont{Johannes and Mazin}(2008)}]{PhysRevB.77.165135}
\bibinfo{author}{\bibfnamefont{M.~D.} \bibnamefont{Johannes}} \bibnamefont{and}
  \bibinfo{author}{\bibfnamefont{I.~I.} \bibnamefont{Mazin}},
  {``}\bibinfo{title}{Fermi surface nesting and the origin of charge density
  waves in metals},{''} \bibinfo{journal}{Phys. Rev. B}
  \textbf{\bibinfo{volume}{77}}, \bibinfo{pages}{165135}
  (\bibinfo{year}{2008}).

\bibitem[{\citenamefont{de~Groot et~al.}(1983)\citenamefont{de~Groot, Mueller,
  van Engen, and Buschow}}]{first_half_met1983}
\bibinfo{author}{\bibfnamefont{R.~A.} \bibnamefont{de~Groot}},
  \bibinfo{author}{\bibfnamefont{F.~M.} \bibnamefont{Mueller}},
  \bibinfo{author}{\bibfnamefont{P.~G.} \bibnamefont{van Engen}},
  \bibnamefont{and} \bibinfo{author}{\bibfnamefont{K.~H.~J.}
  \bibnamefont{Buschow}}, {``}\bibinfo{title}{New Class of Materials:
  Half-Metallic Ferromagnets},{''} \bibinfo{journal}{Phys. Rev. Lett.}
  \textbf{\bibinfo{volume}{50}}, \bibinfo{pages}{2024} (\bibinfo{year}{1983}).

\bibitem[{\citenamefont{Katsnelson et~al.}(2008)\citenamefont{Katsnelson,
  Irkhin, Chioncel, Lichtenstein, and de~Groot}}]{half_met_review2008}
\bibinfo{author}{\bibfnamefont{M.~I.} \bibnamefont{Katsnelson}},
  \bibinfo{author}{\bibfnamefont{V.~Y.} \bibnamefont{Irkhin}},
  \bibinfo{author}{\bibfnamefont{L.}~\bibnamefont{Chioncel}},
  \bibinfo{author}{\bibfnamefont{A.~I.} \bibnamefont{Lichtenstein}},
  \bibnamefont{and} \bibinfo{author}{\bibfnamefont{R.~A.}
  \bibnamefont{de~Groot}}, {``}\bibinfo{title}{Half-metallic ferromagnets: From
  band structure to many-body effects},{''} \bibinfo{journal}{Rev. Mod. Phys.}
  \textbf{\bibinfo{volume}{80}}, \bibinfo{pages}{315} (\bibinfo{year}{2008}).

\bibitem[{\citenamefont{Eschrig}(2015)}]{sc_half_met_eshrig2015}
\bibinfo{author}{\bibfnamefont{M.}~\bibnamefont{Eschrig}},
  {``}\bibinfo{title}{Spin-polarized supercurrents for spintronics: a review of
  current progress},{''} \bibinfo{journal}{Rep. Prog. Phys.}
  \textbf{\bibinfo{volume}{78}}, \bibinfo{pages}{104501}
  (\bibinfo{year}{2015}).

\bibitem[{\citenamefont{Zhong et~al.}(2017)\citenamefont{Zhong, Seyler,
  Linpeng, Cheng, Sivadas, Huang, Schmidgall, Taniguchi, Watanabe, McGuire
  et~al.}}]{Zhonge1603113}
\bibinfo{author}{\bibfnamefont{D.}~\bibnamefont{Zhong}},
  \bibinfo{author}{\bibfnamefont{K.~L.} \bibnamefont{Seyler}},
  \bibinfo{author}{\bibfnamefont{X.}~\bibnamefont{Linpeng}},
  \bibinfo{author}{\bibfnamefont{R.}~\bibnamefont{Cheng}},
  \bibinfo{author}{\bibfnamefont{N.}~\bibnamefont{Sivadas}},
  \bibinfo{author}{\bibfnamefont{B.}~\bibnamefont{Huang}},
  \bibinfo{author}{\bibfnamefont{E.}~\bibnamefont{Schmidgall}},
  \bibinfo{author}{\bibfnamefont{T.}~\bibnamefont{Taniguchi}},
  \bibinfo{author}{\bibfnamefont{K.}~\bibnamefont{Watanabe}},
  \bibinfo{author}{\bibfnamefont{M.~A.} \bibnamefont{McGuire}},
  \bibnamefont{et~al.}, {``}\bibinfo{title}{Van der Waals engineering of
  ferromagnetic semiconductor heterostructures for spin and valleytronics},{''}
  \bibinfo{journal}{Sci. Adv.} \textbf{\bibinfo{volume}{3}}
  (\bibinfo{year}{2017}).

\bibitem[{\citenamefont{Korshunov and Eremin}(2008)}]{PhysRevB.78.140509}
\bibinfo{author}{\bibfnamefont{M.~M.} \bibnamefont{Korshunov}}
  \bibnamefont{and} \bibinfo{author}{\bibfnamefont{I.}~\bibnamefont{Eremin}},
  {``}\bibinfo{title}{Theory of magnetic excitations in iron-based layered
  superconductors},{''} \bibinfo{journal}{Phys. Rev. B}
  \textbf{\bibinfo{volume}{78}}, \bibinfo{pages}{140509}
  (\bibinfo{year}{2008}).

\bibitem[{\citenamefont{Nandkishore
  et~al.}(2012{\natexlab{b}})\citenamefont{Nandkishore, Levitov, and
  Chubukov}}]{NatureSupercondNesting2012}
\bibinfo{author}{\bibfnamefont{R.}~\bibnamefont{Nandkishore}},
  \bibinfo{author}{\bibfnamefont{L.~S.} \bibnamefont{Levitov}},
  \bibnamefont{and} \bibinfo{author}{\bibfnamefont{A.~V.}
  \bibnamefont{Chubukov}}, {``}\bibinfo{title}{Chiral superconductivity from
  repulsive interactions in doped graphene},{''} \bibinfo{journal}{Nat. Phys.}
  \textbf{\bibinfo{volume}{8}}, \bibinfo{pages}{158}
  (\bibinfo{year}{2012}{\natexlab{b}}).

\bibitem[{\citenamefont{Fine}(2007)}]{fine_neutrons2007}
\bibinfo{author}{\bibfnamefont{B.~V.} \bibnamefont{Fine}},
  {``}\bibinfo{title}{Magnetic vortices instead of stripes: Another
  interpretation of magnetic neutron scattering in lanthanum cuprates},{''}
  \bibinfo{journal}{Phys. Rev. B} \textbf{\bibinfo{volume}{75}},
  \bibinfo{pages}{060504} (\bibinfo{year}{2007}).

\bibitem[{\citenamefont{Egami et~al.}(2010{\natexlab{a}})\citenamefont{Egami,
  Fine, Parshall, Subedi, and Singh}}]{fine_review2010}
\bibinfo{author}{\bibfnamefont{T.}~\bibnamefont{Egami}},
  \bibinfo{author}{\bibfnamefont{B.}~\bibnamefont{Fine}},
  \bibinfo{author}{\bibfnamefont{D.}~\bibnamefont{Parshall}},
  \bibinfo{author}{\bibfnamefont{A.}~\bibnamefont{Subedi}}, \bibnamefont{and}
  \bibinfo{author}{\bibfnamefont{D.}~\bibnamefont{Singh}},
  {``}\bibinfo{title}{Spin-lattice coupling and superconductivity in Fe
  pnictides},{''} \bibinfo{journal}{Adv. Condens. Matter Phys.}
  \textbf{\bibinfo{volume}{2010}} (\bibinfo{year}{2010}{\natexlab{a}}).

\bibitem[{\citenamefont{Egami et~al.}(2010{\natexlab{b}})\citenamefont{Egami,
  Fine, Singh, Parshall, de~la Cruz, and Dai}}]{egami_physC2010}
\bibinfo{author}{\bibfnamefont{T.}~\bibnamefont{Egami}},
  \bibinfo{author}{\bibfnamefont{B.}~\bibnamefont{Fine}},
  \bibinfo{author}{\bibfnamefont{D.}~\bibnamefont{Singh}},
  \bibinfo{author}{\bibfnamefont{D.}~\bibnamefont{Parshall}},
  \bibinfo{author}{\bibfnamefont{C.}~\bibnamefont{de~la Cruz}},
  \bibnamefont{and} \bibinfo{author}{\bibfnamefont{P.}~\bibnamefont{Dai}},
  {``}\bibinfo{title}{Spin–lattice coupling in iron-pnictide
  superconductors},{''} \bibinfo{journal}{Physica C: Superconductivity and its
  Applications} \textbf{\bibinfo{volume}{470}}, \bibinfo{pages}{S294 }
  (\bibinfo{year}{2010}{\natexlab{b}}), \bibinfo{note}{proceedings of the 9th
  International Conference on Materials and Mechanisms of Superconductivity}.

\bibitem[{\citenamefont{Lychkovskiy and Fine}(2017)}]{Lychkovskiy2017SpinES}
\bibinfo{author}{\bibfnamefont{O.}~\bibnamefont{Lychkovskiy}} \bibnamefont{and}
  \bibinfo{author}{\bibfnamefont{B.~V.} \bibnamefont{Fine}},
  {``}\bibinfo{title}{Spin excitation spectrum of high-temperature cuprate
  superconductors from finite cluster simulations.},{''}
  \bibinfo{journal}{Journal of physics. Condensed matter : an Institute of
  Physics journal} \textbf{\bibinfo{volume}{30 40}}, \bibinfo{pages}{405801}
  (\bibinfo{year}{2017}).

\bibitem[{\citenamefont{Lee et~al.}(1999)\citenamefont{Lee, Birgeneau, Kastner,
  Endoh, Wakimoto, Yamada, Erwin, Lee, and Shirane}}]{PhysRevB.60.3643}
\bibinfo{author}{\bibfnamefont{Y.~S.} \bibnamefont{Lee}},
  \bibinfo{author}{\bibfnamefont{R.~J.} \bibnamefont{Birgeneau}},
  \bibinfo{author}{\bibfnamefont{M.~A.} \bibnamefont{Kastner}},
  \bibinfo{author}{\bibfnamefont{Y.}~\bibnamefont{Endoh}},
  \bibinfo{author}{\bibfnamefont{S.}~\bibnamefont{Wakimoto}},
  \bibinfo{author}{\bibfnamefont{K.}~\bibnamefont{Yamada}},
  \bibinfo{author}{\bibfnamefont{R.~W.} \bibnamefont{Erwin}},
  \bibinfo{author}{\bibfnamefont{S.-H.} \bibnamefont{Lee}}, \bibnamefont{and}
  \bibinfo{author}{\bibfnamefont{G.}~\bibnamefont{Shirane}},
  {``}\bibinfo{title}{Neutron-scattering study of spin-density wave order in
  the superconducting state of excess-oxygen-doped
  ${\mathrm{La}}_{2}{\mathrm{CuO}}_{4+y}$},{''} \bibinfo{journal}{Phys. Rev. B}
  \textbf{\bibinfo{volume}{60}}, \bibinfo{pages}{3643} (\bibinfo{year}{1999}).

\bibitem[{\citenamefont{Qiu et~al.}(2008)\citenamefont{Qiu, Kofu, Bao, Lee,
  Huang, Yildirim, Copley, Lynn, Wu, Wu et~al.}}]{PhysRevB.78.052508}
\bibinfo{author}{\bibfnamefont{Y.}~\bibnamefont{Qiu}},
  \bibinfo{author}{\bibfnamefont{M.}~\bibnamefont{Kofu}},
  \bibinfo{author}{\bibfnamefont{W.}~\bibnamefont{Bao}},
  \bibinfo{author}{\bibfnamefont{S.-H.} \bibnamefont{Lee}},
  \bibinfo{author}{\bibfnamefont{Q.}~\bibnamefont{Huang}},
  \bibinfo{author}{\bibfnamefont{T.}~\bibnamefont{Yildirim}},
  \bibinfo{author}{\bibfnamefont{J.~R.~D.} \bibnamefont{Copley}},
  \bibinfo{author}{\bibfnamefont{J.~W.} \bibnamefont{Lynn}},
  \bibinfo{author}{\bibfnamefont{T.}~\bibnamefont{Wu}},
  \bibinfo{author}{\bibfnamefont{G.}~\bibnamefont{Wu}}, \bibnamefont{et~al.},
  {``}\bibinfo{title}{Neutron-scattering study of the oxypnictide
  superconductor ${\text{LaFeAsO}}_{0.87}{\text{F}}_{0.13}$},{''}
  \bibinfo{journal}{Phys. Rev. B} \textbf{\bibinfo{volume}{78}},
  \bibinfo{pages}{052508} (\bibinfo{year}{2008}).

\bibitem[{\citenamefont{Mazin and Yakovenko}(1995)}]{Mazin_1995}
\bibinfo{author}{\bibfnamefont{I.~I.} \bibnamefont{Mazin}} \bibnamefont{and}
  \bibinfo{author}{\bibfnamefont{V.~M.} \bibnamefont{Yakovenko}},
  {``}\bibinfo{title}{Neutron Scattering and Superconducting Order Parameter in
  {YBa$_2$Cu$_3$O$_7$}},{''} \bibinfo{journal}{Phys. Rev. Lett.}
  \textbf{\bibinfo{volume}{75}}, \bibinfo{pages}{4134} (\bibinfo{year}{1995}).

\bibitem[{\citenamefont{Qureshi et~al.}(2012)\citenamefont{Qureshi, Steffens,
  Drees, Komarek, Lamago, Sidis, Harnagea, Grafe, Wurmehl, B\"uchner
  et~al.}}]{PhysRevLett.108.117001}
\bibinfo{author}{\bibfnamefont{N.}~\bibnamefont{Qureshi}},
  \bibinfo{author}{\bibfnamefont{P.}~\bibnamefont{Steffens}},
  \bibinfo{author}{\bibfnamefont{Y.}~\bibnamefont{Drees}},
  \bibinfo{author}{\bibfnamefont{A.~C.} \bibnamefont{Komarek}},
  \bibinfo{author}{\bibfnamefont{D.}~\bibnamefont{Lamago}},
  \bibinfo{author}{\bibfnamefont{Y.}~\bibnamefont{Sidis}},
  \bibinfo{author}{\bibfnamefont{L.}~\bibnamefont{Harnagea}},
  \bibinfo{author}{\bibfnamefont{H.-J.} \bibnamefont{Grafe}},
  \bibinfo{author}{\bibfnamefont{S.}~\bibnamefont{Wurmehl}},
  \bibinfo{author}{\bibfnamefont{B.}~\bibnamefont{B\"uchner}},
  \bibnamefont{et~al.}, {``}\bibinfo{title}{Inelastic Neutron-Scattering
  Measurements of Incommensurate Magnetic Excitations on Superconducting LiFeAs
  Single Crystals},{''} \bibinfo{journal}{Phys. Rev. Lett.}
  \textbf{\bibinfo{volume}{108}}, \bibinfo{pages}{117001}
  (\bibinfo{year}{2012}).

\bibitem[{\citenamefont{Mahan}(2000)}]{mahan2013many}
\bibinfo{author}{\bibfnamefont{G.~D.} \bibnamefont{Mahan}},
  \emph{\bibinfo{title}{Many Particle Physics, Third Edition}}
  (\bibinfo{publisher}{Plenum}, \bibinfo{address}{New York},
  \bibinfo{year}{2000}).

\bibitem[{\citenamefont{Furrer et~al.}(2009)\citenamefont{Furrer, Str{\"a}ssel
  et~al.}}]{furrer2009neutron}
\bibinfo{author}{\bibfnamefont{A.}~\bibnamefont{Furrer}},
  \bibinfo{author}{\bibfnamefont{T.}~\bibnamefont{Str{\"a}ssel}},
  \bibnamefont{et~al.}, \emph{\bibinfo{title}{Neutron scattering in condensed
  matter physics}} (\bibinfo{publisher}{World Scientific Publishing Company},
  \bibinfo{year}{2009}).

\end{thebibliography}

\end{document}